\documentclass[letter,twocolumn]{jpsj2}
\usepackage{bm}

\title{Spin-Dependent  Mass Enhancement under Magnetic Field in the Periodic Anderson Model}

\author{Seiichiro \textsc{Onari}$^{1}$\thanks{E-mail address: onari@nuap.nagoya-u.ac.jp}, Hiroshi \textsc{Kontani}$^{2}$ and Yukio \textsc{Tanaka}$^{1}$}

\inst{$^{1}$Department of Applied Physics, Nagoya University, Chikusa, Nagoya 464-8603, Japan.\\
$^{2}$Department of Physics, Nagoya University, Chikusa, Nagoya 464-8603, Japan.\\}

\abst{In order to study the mechanism of the mass 
enhancement in heavy fermion compounds in the presence of 
magnetic field, we study the periodic Anderson model using the fluctuation
exchange approximation. 
The resulting value of the mass enhancement factor $z^{-1}$ can become up to 10, which is
significantly larger than  
that in the single-band Hubbard model.
We show that the difference between the magnitude 
of the mass enhancement factor of up spin (minority spin) electrons $z^{-1}_{\uparrow}$ and 
that of down spin (majority spin) electrons $z^{-1}_{\downarrow}$  
increases by the applied magnetic field $B\parallel z$, 
which is consistent with de Haas-van Alphen measurements for CeCoIn$_5$,
 CeRu$_2$Si$_2$ and CePd$_2$Si$_2$. 
We predict that $z^{-1}_\uparrow>z^{-1}_\downarrow$ in many Ce
compounds, whereas $z^{-1}_\uparrow<z^{-1}_\downarrow$ in Yb compounds.
}

\kword{CeCoIn$_5$, Periodic Anderson model, FLEX}

\begin{document}
\maketitle
Heavy fermion compounds such as CeCoIn$_5$\cite{petrovic} and CeRhIn$_5$\cite{Hegger} have attracted considerable
attention because of their non-Fermi-liquid like electronic properties, where  
these materials have been considered to locate 
close to the antiferromagnetic (AF) quantum critical point (QCP).
The anomalous transport properties specific to the AF QCP  are reported by 
recent experiments of CeCoIn$_5$.  
In the normal state of CeCoIn$_5$, it is clarified that the 
temperature dependence of the resistivity\cite{nakajima} $\rho$ 
and Hall coefficient\cite{nakajima} $R_H$ follows 
$\rho \propto T$ and $R_H\propto 1/T$ below
$30$K.
These behaviors are significantly 
different from the Fermi-liquid behaviors, $i.e.$, 
$\rho\propto T^2$ and $R_H\propto T^0$.
From a theoretical view point, 
we have previously 
derived that $\rho\propto T$ and $R_H\propto 1/T$ 
near the AF QCP by considering the current vertex corrections.
\cite{Onari,Kontani}

It has been also reported that 
these anomalous electronic properties near the AF QCP are sensitive to a magnetic
field. For example, $\rho$, $R_H$ and $\nu$ in CeCoIn$_5$ show a 
nonlinear dependence of the applied 
magnetic field.\cite{Paglione,Bianchi,nakajima} 
Moreover, it
is obtained by de Haas-van Alphen (dHvA) oscillation measurements that
the ratio of the spin-dependent effective mass $\frac{m^*_\uparrow}{m^*_\downarrow}$
reaches $3$ at 15T.\cite{Settai,McCollam}
Therefore,  to 
understand these interesting phenomena, 
theoretical works about heavy fermion compounds in the presence of the 
 magnetic field are needed. 
In particular, it is important to clarify 
a field dependence of the mass enhancement.  

In CeCoIn$_5$, the Fermi surface geometry and renormalized quasiparticle
mass have been obtained by dHvA measurement.\cite{Settai} 
In this compound $4f$-electrons mainly contribute to the density of
state (DOS) at the Fermi level. 
It is known that the applied magnetic field lifts the spin degeneracy, and splits the Fermi surface into majority-spin and minority-spin
surfaces. 
It is obtained that 
effective masses $m^*$ for branch $\beta_2$ of 14th band in
CeCoIn$_5$ at 15T are $m^*=46m_e$ 
for the majority spin and $m^*=154m_e$ for the minority spin,
respectively, where 
$m_e$ denotes the mass of a bare electron.\cite{McCollam} 
The band mass for the $\beta_2$-band is $m_b=1.7m_e$ according to  
the band calculation.\cite{Settai}
Then, the obtained mass enhancement factor $z_\sigma^{-1}=m_\sigma^*/m_b$ 
is 27 for majority spin 
and 90 for minority spin, respectively.
Such a spin-dependent 
mass enhancement is widely observed in heavy fermion compounds such
as CeRu$_2$Si$_2$,\cite{Aoki} CePd$_2$Si$_2$\cite{Sheikin} and
CeIn$_3$\cite{Endo}.

Up to now, electronic properties of heavy fermion systems
under magnetic field have not been studied enough from a 
microscopic theoretical approach.
In the previous studies for the impurity Anderson
model with particle-hole symmetry, it was shown that mass
enhancement factor $(z^{-1}_\uparrow=z^{-1}_\downarrow)$ decreases with magnetic field \cite{hewson}.
Introducing the electron-magnon interaction, Edwards and Green studied the periodic Anderson model (PAM) in
strong magnetic field regime, where number of up-spin $f$-electron
$(n^f_\uparrow)$ is zero.\cite{edwards} However, dHvA measurements are
usually performed under a moderate magnetic field ($\sim 10$T), where
$n^f_\uparrow/n^f_\downarrow \lesssim 1$ is satisfied.
On the other hand, Korbel {\it et al.} studied the Hubbard model using the Gutzwiller approximation and obtained the relationship
$\frac{m^*_\uparrow}{m^*_\downarrow}=\frac{1-n_\uparrow}{1-n_\downarrow}$
for $U=\infty$.\cite{korbel} 
Based on this relationship, 
$\frac{m^*_\uparrow}{m^*_\downarrow}\sim 1.2$ is satisfied for
$\frac{n_\uparrow}{n_\downarrow}\sim 0.9$, which is 
realized under 10T according to the magnetization measurement\cite{tayama}. 
However, a dHvA\cite{McCollam} measurement 
in CeCoIn$_5$ reports $\frac{m^*_\uparrow}{m^*_\downarrow}\sim 3$ under 10T.  
Thus, it is highly desirable to study the origin of the mass enhancement factor 
by taking the strong spin fluctuation near the AF QCP, which is
ignored in the Gutzwiller approximation.
Although Held {\it et al.}\cite{held} and Sakurazawa {\it et al.}\cite{sakurazawa} studied the Hubbard model under  magnetic field using the 
dynamical mean field (DMFT) method and the fluctuation exchange (FLEX) approximation, respectively, 
they did not report spin-dependent mass enhancement.

The purpose of this paper is to study spin-dependent mass
enhancements under magnetic field in 
the PAM. We study the two-dimensional (2D) and three-dimensional (3D)
PAM based on the 
FLEX approximation.
In the PAM, the obtained mass enhancement factor is much larger than
that in the Hubbard model, because of the 
hybridization between $f$-electrons and conduction electrons. 
In both 2D and 3D systems, the mass enhancements of 
up and down
spin quasi-particles 
differ considerably under magnetic field,
which is consistent with 
dHvA measurements in CeCoIn$_5$.\cite{McCollam}. 
We also analyze the mass enhancement factors using the DMFT in the
strong correlation regime, and derived a general 
relationship $\frac{z_\uparrow}{z_\downarrow}\sim
\frac{\rho_\uparrow(0)}{\rho_\downarrow(0)}$, 
where $z_{\sigma}$ and $\rho_{\sigma}(0)$ are 
renormalization factor and density of states at Fermi energy 
with $\sigma=\uparrow(\downarrow)$. 

We start with the PAM,
\begin{eqnarray}
{\cal
 H}&=&\sum_{\bm{k},\sigma}^{\rm 3D}\left[(\epsilon^f+B\sigma)f^\dagger_{\bm{k}\sigma}f_{\bm{k}\sigma}+(\epsilon^c_{\bm{k}}+B\sigma)c^\dagger_{\bm{k}\sigma}c_{\bm{k}\sigma}\right.\nonumber\\
&+&\left.\!\!\!\!\!\!\!\!V(f^\dagger_{\bm{k}\sigma}c_{\bm{k}\sigma}\!+\!c^\dagger_{\bm{k}\sigma}f_{\bm{k}\sigma})\right]\!\!+\!\frac{U}{N}\!\!\sum_{\bm{k,k',q}}\!\!f^\dagger_{\bm{k}\uparrow}f_{\bm{k+q}\uparrow}f^\dagger_{\bm{k'}\downarrow}f_{\bm{k'-q}\downarrow},
\end{eqnarray}
\begin{eqnarray}
\epsilon^c_{\bm{k}}=-2 t_1(\cos(k_x)+\cos(k_y))-4
 t_2\cos(k_x)\cos(k_y)\nonumber\\
-2t_3(\cos(2k_x)+\cos(2k_y))-2t_z\cos(k_z),
\end{eqnarray}
where $f_{\bm{k}\sigma}$ $(f^\dagger_{\bm{k}\sigma})$ is an annihilation
(creation) operator for an $f$-electron, and $c_{\bm{k}\sigma}$
$(c^\dagger_{\bm{k}\sigma})$ is that for a conduction electron, respectively. 
$\epsilon^f$ and $\epsilon^c_{\bm k}$ are the energy of an 
$f$-electron and the dispersion of a conduction electron, respectively. 
$B\sigma$ represents the Zeeman energy, where $B$ is magnetic field and $\sigma=1(-1)$
corresponds to the $\uparrow$-($\downarrow$-) spin state.
Here, we take $g$-factor $g$ and Bohr magneton $\mu_B$ as unity $(g\mu_B=1)$. We consider a stacked square lattice with the Coulomb repulsion $U$ for
$f$-electrons, where the
hybridization is denoted by $V$ and the intralayer hopping
$t_1$, $t_2$, $t_3$ and the interlayer hopping $t_z$ for conduction electrons.
For $t_{z}=0$, the above model is reduced to the 2D PAM.  

Hereafter, we take $t_1=1$ as a unit of energy.
We apply the FLEX
approximation\cite{bickers1,bickers2,koikegami,kontani} where the
Green's function, the self-energy and the 
susceptibility are obtained self-consistently.
The FLEX approximation belongs to the "conserving approximations"
formulated by Baym and Kadanoff\cite{baym-kadanoff,baym}.
From Dyson equation, we obtain Green's function
$G^f_\sigma(k)$ for the $f$-electron and
$G^c_\sigma(k)$ for the conduction electron,
\begin{eqnarray}
G^f_\sigma(k)&=&\frac{1}{i\epsilon_n+\mu-\epsilon^f-B\sigma-\Sigma_\sigma(k)-\frac{V^2}{i\epsilon_n+\mu-\epsilon^c_{\bm{k}}-B\sigma}},\label{gf}\\
G^c_\sigma(k)&=&\frac{1}{i\epsilon_n+\mu-\epsilon^c_{\bm{k}}-B\sigma-\frac{V^2}{i\epsilon_n+\mu-\epsilon^f-B\sigma-\Sigma_\sigma(k)}},
\end{eqnarray}
respectively, where $\Sigma_\sigma(k)$ is the self-energy of
$f$-electrons.
Here and hereafter, $k\equiv(i\epsilon_n,\bm{k})$ 
where $\epsilon_n=(2n+1)\pi T$ is the Matsubara frequency. 
The spin susceptibilities
$\chi_{\sigma,\sigma'}(q)$ and the self-energy $\Sigma_\sigma(k)$ in the
FLEX approximation under magnetic field are formulated in Ref. \cite{sakurazawa}.

Mass enhance factor $z^{-1}_{\bm{k}\sigma}$ is
\begin{equation}
z^{-1}_{\bm{k}\sigma}=1-\left.\frac{\partial{\rm Re}\Sigma_\sigma(\omega,\bm{k})}{\partial \omega}\right|_{\omega=0}+\frac{V^2}{(\mu-\epsilon^c_{\bm{k}}-B\sigma)^2},\label{z}
\end{equation}
where the self-energy $\Sigma_\sigma(\omega,\bm{k})$ 
represented in the real-frequency is obtained by the 
analytic continuation of $\Sigma_\sigma(i\epsilon_n,\bm{k})$ using the
Pade approximation. 
It should be noted that the relationship 
$V^2/(\mu-\epsilon^c_{k_F}-B\sigma)^2\ll 1$ 
is satisfied in actual heavy fermion compounds. 
In the following,  we take $N=N_x\times N_y \times N_z 
=64\times 64\times 64$
$\bm{k}$-point meshes and the Matsubara frequencies $\epsilon_n$ 
from
$-(2N_c-1)\pi T$ to $(2N_c-1)\pi T$ with $N_c=512$, respectively. 


\begin{figure}[htdp]
\begin{center}
\includegraphics[width=8cm]{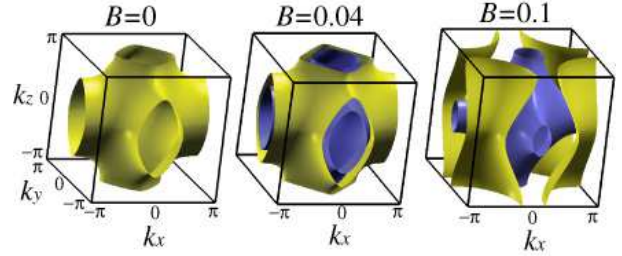}
\caption{(color online) Fermi surface 
in the presence of the magnetic field $B=0$ (left panel), $B=0.04$
 (center) and $B=0.1$ (right) for $T=0.01$, $U=2$ and
 $V=3$. The Fermi surfaces for up-spin and down-spin are
 depicted by the blue (dark) and the yellow (light) sheets, respectively.}
\label{fermi}
\end{center}
\end{figure}

\begin{figure}[htdp]
\begin{center}
\includegraphics[height=50mm]{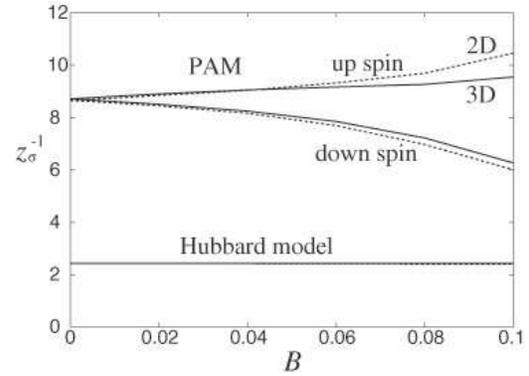}
\caption{Mass enhancement factor $z_\sigma^{-1}$ for an $f$-electron against $B$ for 3D system $t_z=0.8$ 
 (solid line) and for 2D system $t_z=0$ (dotted line) in the PAM with
 $U=2$, $V=3$ and $T=0.01$. $z_\sigma^{-1}$ of up (down) spin in the
 3D Hubbard model with the same Stoner factor $(=0.96)$
for $B=0$ are depicted by solid 
 (dotted) lines. They are almost same lines. 
}
\label{mass}
\end{center}
\end{figure}

\begin{figure}[htdp]
\begin{center}
\includegraphics[height=50mm]{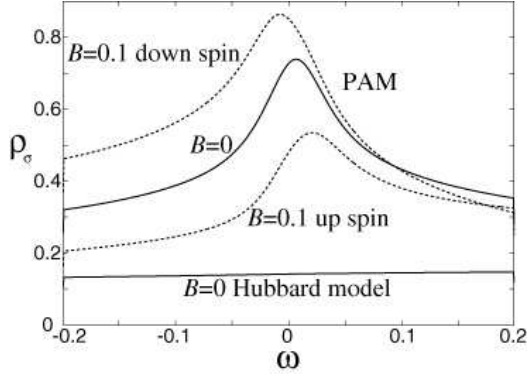}
\caption{DOS of $f$-electron $\rho_\sigma$ against $\omega$ in the 3D PAM for $B=0$ (solid line) and
 for $B=0.1$ (dotted line). The lowest solid line denotes DOS for $B=0$ on the
3D Hubbard model.}
\label{dos}
\end{center}
\end{figure}
In the following, we choose the value of the 
filling of $f$-electrons 
and conduction electrons 
as $n^f=0.8$ $(n^f=n^f_\uparrow+n^f_\downarrow)$, 
and $n^c=0.1$ $(n^c=n^c_\uparrow+n^c_\downarrow)$, respectively, 
by adjusting $\mu$ and $\epsilon^f$. The total electron density is $n=n^f+n^c$.
Here, the up(down) spin corresponds to minority(majority) spin. 
In the actual Ce compounds $n^f\lesssim1$ is satisfied. 
We put the hopping integrals as $t_1=1$, $t_2=-1/6$, 
$t_3=1/5$ and $t_z=0.8$($t_{z}=0)$ 
for three-dimensional (two-dimensional) case. 
 Hereafter in the PAM, we fix parameters $U=2$ and $V=3$.
Strong AF fluctuation with $\bm{Q}=(\pi,\pi,\pi)$
(3D) and $\bm{Q}=(\pi,\pi)$ (2D) is realized for these parameters.
In Fig. \ref{fermi}, we show the calculated 
Fermi surface in the PAM under the magnetic field $B=0$,
$B=0.04$ and $B=0.1$ for $T=0.01$, where the yellow
(light) and
blue (dark) sheets denote the Fermi surfaces for down and
up spin electrons, respectively. 
This Fermi surface resembles that of the 14th band
in CeCoIn$_5$.\cite{shishido} 

The mass enhancement factor $z_{\sigma}^{-1}$ for $f$-electrons, which 
is the averaged value of $(z_{\bm{k}\sigma})^{-1}$ over the 
Fermi surface, is shown in Fig. \ref{mass}. 
$z_{\sigma}^{-1}$
of up- and down-spin electrons at $T=0.01$ in the 3D and 2D PAMs are
shown by full lines and dotted lines, respectively.
In the 2D and 3D PAMs, we see that the 
difference between $z_\uparrow^{-1}$ and $z_\downarrow^{-1}$ increases
with $B$. 
It is  also shown that $z_\sigma^{-1}$ of the minority-spin $(\uparrow)$
electrons is larger than that of the majority-spin $(\downarrow)$ electrons. 
 These results in the PAM are consistent with the experimental data 
of dHvA\cite{McCollam} in CeCoIn$_5$.
In Fig. \ref{mass}, we also plot the mass enhancement factors for up
(solid line) and
down (dotted line) spin electrons, respectively, in the 3D Hubbard
model. In this model, spin dependence of mass enhancement factors is not
visible ($z^{-1}_\uparrow\simeq z^{-1}_\downarrow$).
In order to realize the same value of the 
Storner factor ($=0.96$) in the 3D Hubbard model, we choose the filling $n=0.9$
and $U=4.6$.
Thus, the distance from the AF QCP 
is considered to be the same in both models. 
The resulting values of $z_{\sigma}^{-1}$ in the Hubbard
model are less than  3 and they are  much smaller 
than those in the PAM $(\sim 10)$. 

In the PAM, the band width without renormalization $W$ is estimated by
$W\sim \frac{2V^2}{D}\sim 2$, where $D\sim 10$ is the band width of the
conduction band. Then, $U/W\sim 1$ and $T/W\sim 0.005$ are satisfied in
the PAM. We also calculate the mass enhancement in
the Hubbard model with the same $U/W\sim 1$ and $T/W\sim 0.005$ as in the PAM.
We obtain $z_\uparrow^{-1}\sim 4.5$ for $W\sim 8$, $U=8$ and
$T=0.04$ in the Hubbard model, which is smaller than $z_\uparrow^{-1}\sim 10$ in the PAM.

Here, the magnitude of the magnetic field  $B=0.1$ in the PAM 
corresponds to the situation where 
the magnetic field $H\sim35$T 
is applied to electrons 
of the 14th band of CeCoIn$_5$. 
The reason is as follows. 
The $W$ in CeCoIn$_5$ estimated from the
LDA is $\sim 1000$K.
The Zeeman energy for Ce$^{3+}$ is given by
$(6/7)M\mu_BH$, where $6/7$ is the $g$-value and $M=5/2$ is spin of
Kramers doublet.\cite{sakurazawa}  Then, we can estimate the magnetic
field $H=\frac{0.1}{\frac{6}{7}\frac{5}{2}\mu_B}\times \frac{1000}{2}{\rm K}\sim 35$T for $B=0.1$.

To understand  the detailed behavior of the mass enhancement mechanism, 
we show the DOS $\rho_\sigma(\omega)$ of $f$-electrons, which is
defined by
$\rho_\sigma(\omega)=\sum_{\bm{k}}\rho_{\sigma\bm{k}}(\omega)=\sum_{\bm{k}}\frac{-1}{\pi}{\rm
Im}G_\sigma(\omega+i\delta,\bm{k})$. In Fig. \ref{dos}, we see that
DOS in the PAM is larger than that in the Hubbard model with the same
Storner factor $(=0.96)$.
The resulting DOS of down (majority) spin electrons
is larger than that of up (minority) spin electrons around $\omega=0$ (corresponding to Fermi
energy) under the magnetic field $B=0.1$. 
These  results originate from the fact that the quasi-particle dumping ($|{\rm Im}\Sigma|$) at the Fermi energy for up spin is larger
than that for down spin.

In order to understand the numerical results given by the FLEX approximation,
 we analyze the relationship between the self-energy and
 the resulting DOS based on the infinite-order perturbation theory. 
 For this purpose, we employ the DMFT scheme where
momentum dependence of the self-energy is neglected.
Both in the PAM and the Hubbard model, up spin electrons interact only with
down spin ones.
Then, the resulting self-energy can be expressed as,
\begin{eqnarray}
\Sigma_{\sigma}(i\epsilon_j)=\sum_{n=0}^\infty\sum_{\Gamma_n}\sum_{\{\epsilon_i\},\{\epsilon'_i\}}
 U^{n+1}a(\Gamma_{n+1},n+1,\epsilon_j,\{\epsilon_i\},\{\epsilon'_i\})\nonumber\\
\times G_{\sigma}(i\epsilon_1)\cdots G_\sigma(i\epsilon_n)
G_{-\sigma}(i\epsilon'_1)\cdots G_{-\sigma}(i\epsilon'_{n+1}),
\end{eqnarray}
where $\Gamma_n$ denotes a Feynman diagram of the $n$th order term,
$a(\Gamma_n,n,\epsilon_j,\{\epsilon_i\},\{\epsilon'_i\})$ is a coefficient of 
the $n$th order term
and $\{\epsilon_i\}(\{\epsilon'_i\})$ 
is defined as series of $\epsilon_i$ with $\sigma(-\sigma)$ spin,
respectively. For example, the coefficient of the second order is
$a(\Gamma_2,2,\epsilon_j,\{\epsilon_i\},\{\epsilon'_i\})=-\delta(\epsilon'_2-\epsilon_j+\epsilon_1-\epsilon'_1)$.

Using the spectral representation of the Green's function,
\begin{equation}
G_\sigma(i\epsilon_n)=\int
 d\omega\frac{\rho_{\sigma}(\omega)}{i\epsilon_n-\omega},
\end{equation}
the self-energy is given by,
\begin{eqnarray}
\Sigma_{\sigma}(i\epsilon_j)=\sum_{n=0}^\infty\sum_{\Gamma_n}\sum_{\{\epsilon_i\},\{\epsilon'_i\}}
 \int d\omega_1\cdots d\omega_n d\omega'_1\cdots
 d\omega'_{n+1}\nonumber\\
U^{n+1}a(\Gamma_{n+1},n+1,\epsilon_j,\{\epsilon_i\},\{\epsilon'_i\})\nonumber\\
\times\frac{\rho_{\sigma}(\omega_1)\cdots
\rho_{\sigma}(\omega_n)}{(i\epsilon_{n_1}-\omega_1)\cdots
(i\epsilon_{n_n}-\omega_n)}\frac{\rho_{-\sigma}(\omega'_1)\cdots
\rho_{-\sigma}(\omega'_{n+1})}{(i\epsilon_{n'_1}-\omega'_1)\cdots (i\epsilon_{n'_{n+1}}-\omega'_{n+1})}.
\end{eqnarray}
Here, we assume that 
the contribution around the Fermi energy $(\omega_i=0)$ is dominant in
the integral for $\omega_i$. Then,
the self-energy can be approximately expressed as,
\begin{eqnarray}
\Sigma_{\sigma}(i\epsilon_j)\sim\sum_{n=0}^\infty\sum_{\Gamma_n}\sum_{\{\epsilon_i\},\{\epsilon'_i\}}
 \int d\omega_1\cdots d\omega_n d\omega'_1\cdots
 d\omega'_{n+1}\nonumber\\
U^{n+1}a(\Gamma_{n+1},n+1,\epsilon_j,\{\epsilon_i\},\{\epsilon'_i\})\prod_i\theta(\Omega-|\omega_i|)\theta(\Omega-|\omega'_i|)\nonumber\\
\times\frac{\rho_{\sigma}(0)^n}{(i\epsilon_{n_1}-\omega_1)\cdots
(i\epsilon_{n_n}-\omega_n)}\frac{\rho_{-\sigma}(0)^{n+1}}{(i\epsilon_{n'_1}-\omega'_1)\cdots
(i\epsilon_{n'_{n+1}}-\omega'_{n+1})},
\label{sch}
\end{eqnarray}
where $\Omega\sim W$ is a small cut-off energy.
According to Eq. (\ref{sch}), we obtain
\begin{equation}
\Sigma_{\uparrow}(\omega)\rho_{\uparrow}(0)\sim
 \Sigma_{\downarrow}(\omega)\rho_{\downarrow}(0).
\end{equation}
Therefore, the following relationship is obtained:
\begin{equation}
\frac{z^{-1}_{\uparrow}}{z^{-1}_{\downarrow}}\sim 
 \frac{\rho_{\downarrow}(0)}{\rho_{\uparrow}(0)}\sim\frac{|{\rm Im}\Sigma_\uparrow(0)|}{|{\rm Im}\Sigma_\downarrow(0)|}. 
\label{relation}
\end{equation}
Since all diagrams are taken into account in this analysis, the above
relationship (\ref{relation}) is expected to be
valid for the strongly correlated electron systems.
In the PAM, $z_\uparrow^{-1}>z_\downarrow^{-1}$ is obtained 
from $\rho_\uparrow(0)<\rho_\downarrow(0)$ using eq.  (\ref{relation}).
The reason why the relationship
$\rho_\uparrow(0)<\rho_\downarrow(0)$ is derived can be explained as follows. 
From Eq. (\ref{gf}),
$\rho_\sigma(0)=\rho^{00}(-B\sigma-\Sigma_\sigma(B,\omega=0)+\Sigma_\sigma(0,0))$
is obtained within the DMFT at $T=0$,
where $\rho^{00}(\omega)$ is the DOS of $f$-electrons for $U=0$ and
$B=0$. In the present case, $\epsilon_f+\Sigma(0)>\mu$ is satisfied
$(n\equiv n^f+n^c<2)$. We approximate
$\rho^{00}(\omega)=\sum_{\bm{k}}\delta(\omega+\mu-\epsilon^f-\Sigma_\sigma(0,0)-\frac{V^2}{\mu-\epsilon^c_{\bm{k}}})$
considering that the $\omega$-dependence of DOS of conduction electrons
is weak. $\rho^{00}(\omega)$ increases (decreases) monotonically with
$\omega$ for $\omega\sim 0$ in the case of $n<2$ $(n>2)$.
Thus, the relationship $\rho_\uparrow(0)<\rho_\downarrow(0)$ is derived 
for $B>0$ at sufficiently low temperatures.
Therefore the relationships $z_\uparrow^{-1}>z_\downarrow^{-1}$ and $|{\rm Im}\Sigma_\uparrow(0)|>|{\rm
Im}\Sigma_\downarrow(0)|$ are derived from eq. (\ref{relation}).

\begin{figure}[htdp]
\begin{center}
\includegraphics[height=50mm]{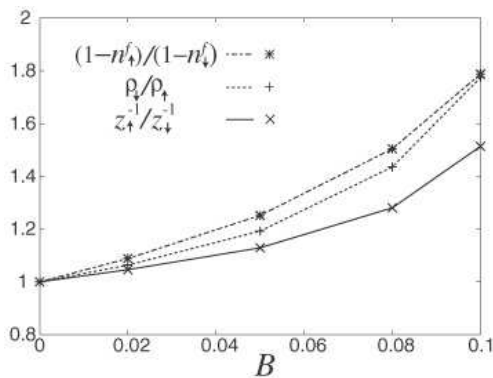}
\caption{$z^{-1}_\uparrow/z^{-1}_\downarrow$ (solid line),
 $\rho_\downarrow(0)/\rho_\uparrow(0)$ (dotted line), and 
 $(1-n_\uparrow^f)/(1-n_\downarrow^f)$ (dashed and dotted line)
 against $B$ in
 the 3D PAM for $T=0.01$.}
\label{comp}
\end{center}
\end{figure}
In the following, we calculate spin-dependent mass enhancements using
the FLEX approximation and confirm
the validity of the analysis in the DMFT.
In Fig. \ref{comp}, $z^{-1}_\uparrow/z^{-1}_\downarrow$ (solid line),
 $\rho_\downarrow(0)/\rho_\uparrow(0)$ (dotted line) are plotted 
as a function of $B$  in the 3D PAM for $T=0.01$. 
We confirm that  $z^{-1}_\uparrow/z^{-1}_\downarrow$ is 
almost proportional to  $\rho_\downarrow(0)/\rho_\uparrow(0)$ in accordance with above relationship (\ref{relation}). In Fig. \ref{comp},
$(1-n^f_\uparrow)/(1-n^f_\downarrow)$ (dashed and dotted line) is also
 shown against $B$, which
represents $z^{-1}_\uparrow/z^{-1}_\downarrow$
in the Gutzwiller approximation 
 for $U=\infty$\cite{korbel}. As shown in Fig. \ref{comp}, $(1-n^f_\uparrow)/(1-n^f_\downarrow)$ is close to
 $z^{-1}_\uparrow/z^{-1}_\downarrow$ obtained by the FLEX approximation in the
 PAM. Thus, $z^{-1}_\uparrow/z^{-1}_\downarrow$ obtained by the FLEX and the 
 Gutzwiller approximations satisfy the relationship (\ref{relation}).

Finally, we discuss discrepancies between our obtained results and experiments.
First one is that the obtained magnitude of $z_\sigma^{-1}$ is much smaller than that of the experimental values with 
$z_\downarrow^{-1}\sim 27$ for majority spin and 
$z_\uparrow^{-1}\sim 90$ for minority spin, respectively, at
$H=15$T.\cite{McCollam}
Second one is that in our calculation the magnitude of $z^{-1}_\uparrow$ increases and that of $z^{-1}_\downarrow$
decreases with magnetic field, while in the experiment
both $z^{-1}_\uparrow$ and $z^{-1}_\downarrow$ decrease with magnetic field. 
Recently, we found that this discrepancy may originate from the absence of vertex corrections 
for the susceptibility in the present calculations.

We have shown that both $\chi_{zz}$ and $\chi_{+-}$ decrease with
 magnetic field when the vertex 
corrections for susceptibility are included in the Hubbard model \cite{onari2}. 
This result leads to the decrease in the magnitude of the mass
enhancement factors of both spins. 

In summary, we have obtained the mass enhancement factor $(z_\sigma^{-1})$ in the 
presence of the magnetic field for the PAM using the FLEX approximation. 
The resulting $z_\uparrow^{-1}$  is almost $10$ and is much larger than
that in the Hubbard model.
The ratio of the mass enhancement factor 
$\frac{z^{-1}_{\uparrow}}{z^{-1}_{\downarrow}}$
increases with the magnetic field. 
Based on the perturbation theory up to infinite order,
we have derived a general relationships,
$i.e.$, $\frac{z^{-1}_{\uparrow}}{z^{-1}_{\downarrow}}\sim
 \frac{\rho_{\downarrow}(0)}{\rho_{\uparrow}(0)}$
by applying the DMFT.
In the case of $n<2$ $(\epsilon^f+\Sigma(0)>\mu)$, which corresponds to Ce compounds, the relationship $z^{-1}_\uparrow >
z^{-1}_\downarrow$ is obtained,
since the DOS of the up-spin electrons at the Fermi energy is smaller than
that of down-spin electrons. This result
is consistent with the experimental data of de Haas-van Alphen\cite{McCollam}
in CeCoIn$_5$. We predict that $z^{-1}_\uparrow <
z^{-1}_\downarrow$ in Yb compounds, which correspond to $n>2$ $(\epsilon^f+\Sigma(0)<\mu)$.

{\bf Acknowledgments}

This work was supported by a Grant-in-Aid for 21st Century COE
``Frontiers of Computational Science''.
Numerical calculations were performed at the supercomputer center, ISSP.

\end{document}